\documentstyle[pra,aps,preprint]{revtex}
\begin{document}
\draft
\title{
Molecular-orbital theory for the stopping
power of atoms in the low velocity regime:
the case of helium in alkali metals.
      }
\author{
\bf Jos\'e J. Dorado and F. Flores. }
\address{
Departamento de F\'\i sica de la Materia Condensada C-XII,
Facultad de Ciencias, \\  Universidad Aut\'onoma de Madrid,
E-28049 Madrid. Spain.
        }
%
\maketitle
\begin {abstract}
  A free-parameter linear-combination-of-atomic-orbitals approach is
presented for analyzing the stopping power of slow ions moving in a
metal. The method is applied to the case of He moving in alkali metals.
Mean stopping powers for He present a good agreement with local density
approximation calculations. Our results show important variations in the
stopping power of channeled atoms with respect to their mean values.
\end{abstract}
\pacs{34.50.Bw , 61.80.Mk }
%
%
\section{Introduction}
The interaction of ions with condensed matter has drawn the
attention of
many researchers from the beginning of this century
\cite{rut:pm21:11} . A
great deal of
work in this field has dealt with the energy loss of swift ions
in
solids. In this regard the work of Bethe \cite{bet:ap5:30} , Fermi
\cite{fer:zf29:27} , Williams \cite{wil:rmp17:45},
and Lindhard \cite{lin:mfm28:54} opened the modern way of calculating
the stopping power of
swift ions in condensed matter. The case of low-velocity projectiles is
much more complicated due to the strong interaction of the moving ion
with the solid. In this case, the projectile is dressed by a number of
electrons that strongly screen the ion-solid interaction. Brandt
\cite{bra:acs:75}
introduced a
Thomas-Fermi statistical model in order to define an effective
ion charge that takes into account how the bound electrons dress the
projectile. Other researchers \cite{rit:jpcs26:65} have developed the
Lindhard approach and
have calculated the stopping power using a linear-response
function. In
recent developments \cite{arn:prl65:90,flo:icpss:91} , the
stopping power for ions
at low and intermediate velocities has been obtained by introducing the
different
electron-loss and -capture processes associated with the
interaction of
the projectile with the target \cite{ech:ssps43:90} .

An important development in the calculation of the stopping power
for
very-low-velocity ions in solids appeared with the application
of the
local density theory to this field.
Echenique, Nieminen, and Ritchie \cite{ech:ssc37:81}
calculated the stopping power for very slow ions moving in a
uniform
electron gas, using well-known techniques in this field. This
approach
has yielded a substantial improvement in the agreement between
experimental data and theoretical calculations. The main
limitation of
this approach, as it has been used in the actual calculations,
is the assumption of having an uniform electron gas
in
the solid. Although some attempts are currently
tried to
improve over this simplification
\cite{bau:prl69:92,gra:pla:92}, it could be convenient to try,
at
the same time, other alternatives that might be appropriate in
the case of having very ionic or covalent solids.

The aim of the work presented in this paper is to apply to the
stopping
power field an approach recently developed for the calculation
of
electronic properties of solids \cite{gol:prb39:89,gar:prb44:91}
. This is a linear-combination-of-atomic-orbitals (LCAO)
approach, whereby the
electronic properties of the solids are calculated from the
localized wave functions of the atoms of the solid. This approach
tries
to emphasize the local chemical properties of the solid and is
deeply
related to the work done by other groups, trying to calculate the
stopping power for ions in solids using the stopping power in the
vapour
target \cite{sab:pra42:90} . The advantage of these approaches is
related
to the non-uniformity of the target, since a
local-density-approximation (LDA) calculation usually
assumes
a uniform electron gas inside a crystal. Thus the long-term aim
of our
approach is first,
     to calculate the stopping power for ions as a function
of the
ion position, in particular, near crystal surfaces; and
     secondly, to take into account the contribution of the
different
atomic
orbitals of the target, mainly those orbitals which have such a
localized size that can not be replaced by an uniform electron
gas.

In this paper we have chosen to analyze the case of helium
interacting
with alkali metals. This is a case in which the interaction of
the
projectile and the target is simple. It is, however, a
complicated system since it  presents a long-range interaction
between
orbitals located at large separations. In these
metals, the local density approach can be expected to be very good;
therefore, we have
chosen it as a strong test to the method we have developed, and
the results obtained give strong support to it.

In the Secs. II and III, we present our model, the formalism used
to solve it, and its application to the case of He in metals.
 In Sec. IV we discuss our results, and in Sec. V we present our
conclusions.


\section{Model and formalism}
\subsection{General formalism}
Our model is an extension of a previous approach to the
calculation of the electronic
properties of the solids using a LCAO method
\cite{gol:prb39:89,gar:prb44:91}.
The basic idea is to
introduce the atomic orbitals $ \psi_{\nu} , \nu =i,\alpha$, $i$
referring to
the crystal site and  $\alpha$ to a particular orbital, and the
orthonormal basis
  $\phi_{\mu}$
\begin{equation}
 \phi_{\mu} = \sum_{\nu} (S^{-1/2})_{\mu,\nu}
\psi_{\nu},
        \label{eq:-1}
\end{equation}
with
\begin{equation}
   S_{\nu \mu} = \langle \psi_{\nu} \mid \psi_{\mu} \rangle,
\end{equation}
obtained using L\"{o}wdin's orthonormalization procedure
\cite{Low}.
Using this new basis, the electron Hamiltonian of a given system
can be
written in the following way:
\begin{eqnarray}
  \hat{H} = & \sum_{\nu,\sigma}
E_{\nu}^{\sigma}\hat{n}_{\nu\sigma} +\sum_{\nu,\mu\neq\nu,\sigma}
T_{\nu\mu}^{\sigma} (c_{\mu\sigma}^{\dagger}c_{\nu\sigma} +
c_{\nu\sigma}^{+}c_{\mu\sigma})
 + \sum_{\nu} U_{\nu}^{(0)} \hat{n}_{\nu\uparrow}
\hat{n}_{\nu\downarrow}   \nonumber \\
& + \frac{1}{2} \sum_{\nu,\mu\neq\nu,\sigma} [ J_{\nu\mu}^{(0)}
\hat{n}_{\mu\sigma} \hat{n}_{\nu-\sigma}
+ (J_{\nu\mu}^{(0)} - J_{x,\nu\mu}^{(0)} + J_{\nu\mu}^{(0)}
S_{\nu\mu}^{2}) \hat{n}_{\mu\sigma} \hat{n}_{\nu\sigma} ]
     \label{eq:1}
\end{eqnarray}
with the operators $c_{\nu\sigma}^{\dagger}$ and $c_{\nu\sigma}$
related to the
orthonormalized wave functions $\phi_{\nu}$. The
different terms in Eq. (\ref{eq:1}) are analyzed in Refs.
\cite{gol:prb39:89} and \cite{gar:prb44:91} ; here we only
comment how to introduce the many-body terms of hamiltonian
(\ref{eq:1})
in a one-body Hamiltonian by means of a Slater-like
potential. This implies replacing Eq. (\ref{eq:1}) by the effective
Hamiltonian:
\begin{equation}
 \hat{H}_{{\rm eff}} =  \sum_{\nu,\sigma}
\tilde{E}_{\nu}^{\sigma}  \hat{n}_{\nu\sigma}
+ \sum_{\nu,\mu\neq\nu,\sigma} T_{\nu\mu}^{\sigma}
(c_{\mu\sigma}^{+}c_{\nu\sigma} + c_{\nu\sigma}^{+}c_{\mu\sigma})
       \label{eq:1a}
\end{equation}
where
\begin{eqnarray}
 \tilde{E}_{\nu}^{\sigma} & = & E_{\nu}^{\sigma} + U_{\nu}^{(0)}
\langle \hat{n}_{\nu -\sigma} \rangle  \nonumber \\
  &  &
    + \sum_{nu,\mu\neq\nu}
     J_{\nu\mu}^{(0)}
             \langle \hat{n}_{\mu  - \sigma} \rangle
    + \sum_{nu,\mu\neq\nu}
         ( J_{\nu\mu}^{(0)}  - J_{x, \nu \mu}^{0}
          +J_{\nu\mu}^{(o)} S_{\nu \mu}^{2}
          +V_{\nu \mu}^{{\rm x,c}} )
          \langle \hat{n}_{\mu  \sigma} \rangle
\end{eqnarray}
$ V_{\mu}^{x,c} $ is the exchange and correlation potential
\cite{gar:prb44:91}
associated with the many-body terms of Eq. (\ref{eq:1}).

We start from Hamiltonian  (\ref{eq:1a}) and assume that its solution can
be obtained in the static limit $ v \rightarrow 0 $, for the case
of an atom moving inside a crystal (see Fig. 1). In our model,
the different parameters of Hamiltonian (\ref{eq:1a}), as well as its
static solution, are calculated for a geometrical configuration,
at each position of the external atom inside the crystal.

To proceed further, we assume that, due to the atomic motion,
there is a time dependence of hamiltonian (\ref{eq:1a}) through the ion
velocity. This implies introducing a quasiadiabatic Hamiltonian,
$\hat{H}_{{\rm eff}}(t)$, with the different parameters,
$E_{\nu}^{\sigma}$ and $T_{\nu\mu}$, having an explicit, but
slowly, time dependence.

In order to calculate the stopping power at a given time and
atomic position, the static solution of hamiltonian
$\hat{H}_{{\rm eff}}$ is introduced. This implies writing
\begin{equation}
   \hat{H}_{eff} \mid n \rangle = E_{n} \mid n \rangle .
\end{equation}

Then, the stopping power (written as a function of the local time
t, defining the projectile position) is given by the following
equation
\cite{sol:td:85}
\begin{equation}
 \frac{dE}{dt} = -2 {\rm Re} \sum_{n} \int_{-\infty}^{t} dt{'}
\frac{e^{-iw_{n0}(t-t^{'})}}{ w_{n0}} \langle 0 \left|
\frac{d\hat{H}_{ {\rm eff}}(t)}{dt} \right| n
        \rangle \langle n \left|
\frac{d\hat{H}_{{\rm eff}}(t^{'})}{dt^{'}} \right| 0 \rangle .
           \label{eq:5}
\end{equation}
[We are using atomic units ($\hbar = m = e^{-} =1$).]
Equation (\ref{eq:5}) is only valid in the quasiadiabatic limit,
with the ion
velocity going to zero. Notice that in Eq. (\ref{eq:5}), the eigenstates
$\mid n \rangle$ correspond to the full Hamiltonian $\hat{H}_{{\rm
eff}}$, including
the external ion, at the final time t. This
approximation is obviously only appropriate for $v \rightarrow 0$.

Equation (\ref{eq:5}) can be further modified by noting that
the dependence of
$\hat{H}_{{\rm eff}}$  with t appears through the coordinate ${\bf R}
= {\bf R}_{0} + {\bf v} t$, of the external atom. Thus we
write $ \frac{d \hat{H}_{{\rm eff}}(t)}{dt}  =
 ( {\bf v} \cdot  {\bf \nabla} ) \hat{H}_{{\rm eff}} ( {\bf R})$, and
introduce
the Fourier-transform $ \hat{H}_{{\rm eff}}( {\bf q})$ of
$\hat{H}_{{\rm eff}}( {\bf R})$. This yields
\begin{eqnarray}
 \frac{dE}{dt} & = &
          -2 {\rm Re} \sum_{n}
          \int \frac{ d {\bf q} }{ ( 2 \pi) ^{3} }
               \frac{ d {\bf q}' }{ ( 2 \pi) ^{3} }
          \int_{-\infty}^{t}
          \frac{
              e^{-i w_{n0} ( t-t')}}
                { w_{n0} }
           ({\bf q} \cdot {\bf v} )
           ({\bf q'} \cdot {\bf v} )     \nonumber \\
       &  &
  \times   e^{i {\bf q} \cdot ( {\bf R}_{0} + {\bf v} t) }
            e^{- i {\bf q}' \cdot ( {\bf R}_{0} + {\bf v} t') }
          \langle 0 \mid \hat{H}_{eff}( {\bf q} ) \mid n \rangle
     \langle n \mid \hat{H}_{eff}( {\bf q}') \mid 0 \rangle .
            \label{eq:6}
\end{eqnarray}
This equation can be easily integrated on t$^{'}$. Moreover, we
introduce the one-electron eigenfunctions and eigenvalues,
$\mid {\bf k} \rangle$,
 $\varepsilon_{k}$ of
Hamiltonian $\hat{H}_{{\rm eff}}$ in Eq. (\ref{eq:6}) to define
$\mid  n \rangle$ and $w_{n0}$.

These steps yield the following results:
\begin{eqnarray}
   \frac{dE}{dt} & = &
   4 \pi
   \sum_{k \langle k_{F} , k' \rangle k_{F}}
   \int \frac{ d {\bf q} }{(2\pi)^{3}}
   \int \frac{d {\bf q'} }{(2\pi)^{3}}
   \frac{ ( {\bf q} \cdot {\bf v}   )
          ( {\bf q} \ ' \cdot {\bf v} )}
        { w_{kk'}}  \nonumber \nonumber \\
            &  &  \times
     \langle {\bf k}' \mid
            \hat{H}_{{\rm eff}}( {\bf q})
            e^{i {\bf q} \cdot  {\bf R} }
     \mid {\bf k}  \rangle
     \langle {\bf k}  \mid
            \hat{H}_{{\rm eff}}( {\bf q})
            e^{-i {\bf q}' \cdot  {\bf R} }
     \mid {\bf k}' \rangle \nonumber \\
            &  & \times
     \delta( w_{kk'} + {\bf q}  \cdot {\bf v} ) ,
       \label{eq:7}
\end{eqnarray}
where the spin has been added up and
$w_{kk'} = \varepsilon_{k'} - \varepsilon_{k}$ . Note that
$\mid {\bf k} \rangle$ and $\varepsilon_{k}$  are the eigenfunctions and
eigenvalues of
the total Hamiltonian, $\hat{H}_{{\rm eff}}( {\bf R})$ , with the
external ion
included. One should remember, however, that
$\mid {\bf k} \rangle$  and  $\mid {\bf k}' \rangle$ are not
eigenfunctions of $\hat{H}_{{\rm eff}}( {\bf q}):$
%
\begin{equation}
      \hat{H}_{{\rm eff}}( {\bf q}) =
      \int d {\bf R}' \
      e^{-i {\bf q} \  \cdot  {\bf R} \ ' }
      \hat{H}_{{\rm eff}}( {\bf R} \ ' ) .
       \label{eq:7a}
\end{equation}

It is of interest to make contact between Eq. (\ref{eq:7})  and
the linear-response
theory. In this case, the total Hamiltonian is written as the sum
of the
unperturbed hamiltonian $\hat{H}_{0}$ and a perturbation
$\hat{H}_{{\rm pert}} = \hat{V}$. Then, Eq. (\ref{eq:7}) can be
transformed by taking for
$\mid {\bf k} \rangle$  and $\mid {\bf k}' \rangle$ ,
 the eigenfunctions of  $\hat{H}_{0}$; moreover, the
perturbation $\hat{V}$, can be written as follows:
\begin{equation}
  \hat{V}  =
   \int d {\bf r} \
    \frac{Z}{
         \mid {\bf R} - {\bf r} \mid
            }
        \hat{\rho}( {\bf r}) ,
\end{equation}
where $Z$ is the external ion charge and ${\bf R}$ its position.
Then, the
power loss is given by the following equation (linear theory):
\begin{eqnarray}
   \frac{dE}{dt}  & = &
   4 \pi
   \sum_{k \langle k_{F} , k' \rangle k_{F}}
   \int \frac{ d {\bf q}  }{(2\pi)^{3}}
   \int \frac{ d {\bf q}' }{(2\pi)^{3}}
   \left( \frac{4\pi Z}{q^{2}} \right)
   \left( \frac{4\pi Z}{q'^{2}} \right)
          ( {\bf q} \cdot {\bf v} )
         \nonumber \\
    & & \times
    e^{i( {\bf q}- {\bf q}' ) \cdot  {\bf R} }
     \langle {\bf k}' \mid
            \rho ^{+} ( {\bf q})
     \mid {\bf k} \rangle
     \langle {\bf k}  \mid
            \rho( {\bf q}')
     \mid{\bf k}' \rangle
     \delta( w_{kk'} + {\bf q}  \cdot {\bf v} )
           \label{eq:9a}
\end{eqnarray}
or, equivalently,
\begin{eqnarray}
   \frac{dE}{dt} & = &
   2 \
   \int \frac{ d {\bf q}  }{(2\pi)^{3}}
   \int        d {\bf q'}
   \left( \frac{4\pi Z}{q^{2}} \right)
   \left( \frac{4\pi Z}{q'^{2}} \right)
          ( {\bf q} \cdot {\bf v} ) \nonumber \\
            &  & \times
    e^{i( {\bf q}- {\bf q}' ) \cdot  {\bf R} }
    {\rm Im} \chi ( {\bf q}, {\bf q}' ;- {\bf q} \cdot {\bf v}),
           \label{eq:9b}
\end{eqnarray}
where
  ${\rm Im} \chi ( {\bf q}, {\bf q}' ;w)$
is the metal polarizability.

For an homogeneous system, only $ {\bf q} = {\bf q}' $ contributes,
and Eq. (\ref{eq:9b}) yields
\begin{equation}
   \frac{dE}{dt}  =
   2 \
   \int \frac{d {\bf q} }{(2\pi)^{3}}
   \left( \frac{4\pi Z}{q^{2}} \right) ^{2}
          ( {\bf q} \cdot {\bf v} )
    {\rm Im} \chi ( {\bf q};- {\bf q} \cdot {\bf v}) ,
\end{equation}
in agreement with other Refs. \cite{lin:mfm28:54,sol:td:85}.
Equation (\ref{eq:7}) is the basic equation giving the stopping power of
the moving
ion, in the low velocity limit, within our LCAO approach. In
Eq. (\ref{eq:7})
the critical quantity to calculate, using the static interaction
between the external charge and the solid, is
$\langle {\bf k} \mid \hat{H}_{{\rm eff}} ( {\bf q} ) \mid {\bf k}' \rangle$.
In
this paper we
shall concentrate on the He case; this provides a simple case in
which to test the method discussed here.
%
\subsection{Static interaction of He with a metal}
In this section, we will present a summary of the main results
discussed
in Ref. \cite{gol:prb39:89}. We shall also extend this discussion in
order to calculate the
matrix elements
$\langle {\bf k} \mid \hat{H}_{{\rm eff}} ( {\bf q} ) \mid {\bf k}' \rangle$,
needed for the calculation of the stopping power.
Following Ref. \cite{gol:prb39:89}, we start by considering the
one-electron
interactions between the He 1$s$ level and a metal band that is
represented in Fig. 2 by a half-occupied $s$ level.
As discussed in Ref. \cite{gol:prb39:89}, there are two
different one-electron
interactions. First, due to the overlap $S$ between the He
1$s$ wave function
and the metal orbital
$(S = \langle \psi_{M} \mid \psi_{{\rm He}} \rangle)$, there is an increase
in the
kinetic energy of the electrons of the system.
This is measured by the following shift of the one-electron
terms:
\begin{equation}
       \delta E_{M}^{(1)} =
            \frac{1}{4} S^{2} ( E_{M}^{0} -E_{{\rm He}}^{0}    )
                      -ST ,
        \label{eq:10a}
\end{equation}
\begin{equation}
       \delta E_{{\rm He}}^{(1)} =
          -  \frac{1}{4} S^{2} ( E_{M}^{0} -E_{{\rm He}}^{0}   )
                      -ST ,
        \label{eq:10b}
\end{equation}
where $T$, the hopping between the two orbitals,
 $\psi_{M} $ and $ \psi_{{\rm He}}$,
is found to be
$- \frac{1}{2} S ( E_{M}^{0} -E_{{\rm He}}^{0})$.
$E_{M}^{0}$ and $E_{{\rm He}}^{0}$ are the metal and He energy levels.
Second, due to the hopping $T$ between the two orbitals we find a
hybridization contribution to the total energy given by the
following shift in
$E_{M}^{0}$ and $E_{{\rm He}}^{0}:$
\begin{equation}
       \delta E_{M}^{(2)} =
            \frac{T^{2}}{ ( E_{M} -E_{{\rm He}}  )      } ,
        \label{eq:11a}
\end{equation}
\begin{equation}
       \delta E_{{\rm He}}^{(2)} =
            - \frac{T^{2}}{ ( E_{M} -E_{{\rm He}}  )      } .
          \label{eq:11b}
\end{equation}
Combining Eqs. (\ref{eq:10a})-(\ref{eq:11b}) ,
we find the following contributions:
\begin{equation}
       \delta E_{M} =
             S^{2} ( E_{M} -E_{{\rm He}}    ),
         \label{eq:11c}
\end{equation}
\begin{equation}
       \delta E_{{\rm He}} =
                       0.
\end{equation}

These shifts in the one-electron levels yield the following
contribution
to the repulsive energy:
\begin{equation}
    \delta V_{{\rm repulsive}}^{{\rm one-electron}}  =
            n_{M}
             S^{2} ( E_{M} -E_{{\rm He}}    ),
      \label{eq:13}
\end{equation}
where $n_{M}$ is the number of electrons in the metal
orbital.

Many-body contributions have also been discussed in Ref.
\cite{gol:prb39:89}. These terms
can be written in a way similar to Eq. (\ref{eq:13});
 in Ref. \cite{gol:prb39:89} it was found
that the total repulsive energy between the metal atom and
He is given by
\begin{equation}
    \delta V_{{\rm repulsive}}^{{\rm one-electron}}  =
            n_{M}
             S^{2} ( E_{M} -E_{{\rm He}}    )
           + n_{M}
             ( -J_{x}^{0} + S^{2} J_{0} )
           + V_{{\rm electrostatic}} ,
         \label{eq:14}
\end{equation}
where $J_{x}^{0}$ is the exchange integral between the metal
and the
He orbitals, $J_{0}$ the Coulomb interaction between the same
orbitals, and $V_{{\rm electrostaric}}$ the electrostatic interaction
between the
total charges of the two atoms. For a He-orbital going like
$ (\frac{\beta ^{3} }{\pi})^{1/2}
  e^{-\beta r }
$
we find that
\begin{equation}
    -J_{x}^{0} +
     V_{{\rm electrostatic}}
          =
          - \frac{3}{8} \beta S^{2} .
\end{equation}
This shows that the repulsive potential can be written as
follows:
\begin{equation}
   V_{{\rm repulsive}} =
            n_{M}
             S^{2} ( E_{M} -E_{{\rm He}}
             - \frac{3}{8} \beta + J_{0} ).
      \label{eq:15b}
\end{equation}

In our actual problem we are interested in calculating
$\langle {\bf k}' \mid \hat{H}_{{\rm eff}} ( {\bf R} ) \mid {\bf k}
\rangle$
, the matrix element
of the total Hamiltonian between the one-electron states
   $\mid {\bf k} \rangle$
. We will show how Eq. (\ref{eq:15b}) can be related to
 $\langle {\bf k}' \mid \hat{H}_{{\rm eff}} ( {\bf R} ) \mid {\bf k}
 \rangle $
. To this end, we
start by discussing the solution of the total Hamiltonian
(crystal plus the external
atom)
within a one-electron approximation. The solution of this
Hamiltonian $\hat{H}$ is given by
\begin{equation}
   \psi =
     \sum_{k}
       c_{k} \psi_{k} +
       c_{{\rm He}} \psi_{{\rm He}} ,
        \label{eq:16}
\end{equation}
where
       $\psi_{k}$
are the eigenfunctions of the crystal Hamiltonian, $\hat{H}_{0}
$, and
   $\psi_{{\rm He}}$
the 1s orbital of He. In writing Eq. (\ref{eq:16}) ,
we assume that the total
Hamiltonian (in our one-electron approximation) is given by
    $\hat{H} = \hat{H}_{0} + \hat{V}_{{\rm He}}$
, where $\hat{V}_{{\rm He}}$
defines the one-electron potential created by the atom.
The eigenvalues and
the eigenfunctions of $\hat{H}$ are given by the secular
equation
\begin{equation}
  {\rm det} \mid \langle \psi_{i} \mid -E + \hat{H} \mid \psi_{j}
         \rangle
       \mid = 0.
\end{equation}

Now, we follow Ref. \cite{gol:prb39:89} and introduce the
orthonormalized
wave functions [as done in Eq. (\ref{eq:-1}) for the basis
$\psi_{ \nu }$]
\begin{equation}
  \phi_{i} =
      \sum_{i'}
      ( S^{-1/2})_{ii'}
     \psi_{i'} ,
       \label{eq:18a}
\end{equation}
with
\begin{equation}
      S_{k{\rm He}} = \langle \psi_{k} \mid \psi_{{\rm He}} \rangle,
\end{equation}
and
\begin{equation}
      S_{kk'} = \langle \psi_{k} \mid \psi_{k'} \rangle= 0 .
\end{equation}
Using Eq. (\ref{eq:18a}) we define the following effective
hamiltonian
\begin{equation}
  \hat{H}_{{\rm eff}} =
       S^{-1/2}
       \hat{H}
       S^{-1/2}.
\end{equation}

In Ref. \cite{gol:prb39:89}, the diagonal terms of the effective
Hamiltonian were calculated up to second order in the overlap, a
small parameter used for calculating $S^{-1/2}$ in a
series expansion, while
the off-diagonal terms  were only obtained up to first order.
In our actual problem we need to calculate
$(\hat{H}_{{\rm eff}})_{kk'}$ up to second order in the overlap, the
smallest surviving term of the expansion.

Proceeding in this way, we obtain the following  results:
\begin{equation}
   (\hat{H}_{{\rm eff}})_{k{\rm He}}
        =
      T_{k{\rm He}} =
        - \frac{1}{2} S_{k{\rm He}} ( E_{k}^{0} -E_{{\rm He}}^{0} ) ,
        \label{eq:20a}
\end{equation}
\begin{eqnarray}
   (\hat{H}_{{\rm eff}})_{kk'}=
      T_{kk'} & =   &
           (V_{{\rm He}})_{kk'}
        - \frac{1}{2}
         (   T_{k{\rm He}} S_{{\rm He}k'} +
            T_{k'{\rm He}} S_{{\rm He}k}   )    \nonumber \\
           & &
        + \frac{1}{4}
       ( \frac{E_{k}^{0}+E_{k'}^{0}}{2} - E_{{\rm He}} )
             S_{{\rm He}k'}
             S_{{\rm He}k} ,
          \label{eq:20b}
\end{eqnarray}
where $E_{k}^{0}$ and $E_{{\rm He}}^{0}$ are the k-state and the
atomic levels, respectively.

Equation (\ref{eq:20a}) was already discussed in
Ref. \cite{gol:prb39:89}
, and found to be valid for a very localized
wavefunction like the He 1$s$ level. Equation (\ref{eq:20b})
is the new
equation we are looking for; here
$ (V_{He})_{kk'} $
is associated with the direct perturbation introduced by the He atom
on the metal. This perturbation is basically due to the atomic
Hartree
potential, and to the exchange perturbation created by the He
1$s$ level.

Equations (\ref{eq:20a}) and (\ref{eq:20b}) can be further
approximated by taking
$E_{k}^{0}$, the one-electron $k$ state levels, equal to
$E_{M}^{0}$  a mean level of the metal band (notice that
the He level $E_{{\rm He}}^{0}$ is very deep and that replacing
$E_{k}^{0}$ by $E_{M}^{0}$ is a good approximation) . Then
Eqs. (\ref{eq:20a}) and (\ref{eq:20b}) read
\begin{equation}
      T_{k{\rm He}} =
        - \frac{1}{2} S_{kHe} ( E_{M} - E_{{\rm He}} ) ,
            \label{eq:21a}
\end{equation}
\begin{eqnarray}
      T_{kk'} & = &
           (V_{{\rm He}})_{kk'}
        - \frac{1}{2}
          (  T_{k{\rm He}} S_{{\rm He}k'} +
            T_{k'{\rm He}} S_{{\rm Hek}}  )
        + \frac{1}{4}
       ( E_{M}- E_{{\rm He}} )
             S_{{\rm He}k'}
             S_{{\rm He}k}  \nonumber \\
       & = &
           (V_{{\rm He}})_{kk'}
        + \frac{3}{4}
       ( E_{M}- E_{{\rm He}} )
             S_{{\rm He}k'}
             S_{{\rm He}k} . \label{eq:21b}
\end{eqnarray}
The terms appearing in Eq. (\ref{eq:21b}), that depend
on $T_{k{\rm He}}$
 and $S_{k{\rm He}}$, are equivalent to the ones going like
$(-ST)$
, in Eq. (\ref{eq:10a}), if $T$ is replaced here by
$-\frac{1}{2} S ( E_{M} - E_{{\rm He}} )$
; this shows how the one-electron correction to the metal level
$\frac{3}{4} S^{2} ( E_{M} - E_{{\rm He}} )$
coincides with the one-electron contribution to the off-diagonal
term in
$T_{kk'}$ if $S^{2}$ is replaced by
$S_{k{\rm He}} S_{{\rm He}k'}$.

Returning to Eq. (\ref{eq:21a}), we should comment that
$T_{k{\rm He}}$ is a first order term in the overlap $S_{{\rm
Hek}}$
while $T_{kk'}$
is second order [$(V_{{\rm He}})_{kk'}$ included]. The first order
term $T_{k{\rm He}}$ introduces an
effective second order contribution to
$T_{kk'}$ given by
\begin{equation}
   \frac{T_{k{\rm He}}
         T_{{\rm He}k'}}
        {E_{M}^{0}-E_{{\rm He}}^{0}} .
       \label{eq:22}
\end{equation}

 Combining Eqs. (\ref{eq:21a}) and (\ref{eq:21b}) with Eq.
(\ref{eq:22}) we get the following effective interaction:
\begin{equation}
   T_{kk'} =
         (V_{{\rm He}})_{kk'}
        + (E_{M}^{0}-E_{{\rm He}}^{0})
            S_{k{\rm He}} S_{{\rm He}k'} .
       \label{eq:22a}
\end{equation}
This is the one-electron contribution to the effective hopping
between the crystal wave functions  $\mid {\bf k} \rangle$ and
$\mid {\bf k}' \rangle$, as induced by
the external atom. When the crystal wavefunctions
$\mid {\bf k} \rangle$
 are developed in a local basis
\begin{equation}
    \mid {\bf k} > =
     \sum_{i} c_{i}( {\bf k}) \phi_{i} ,
\end{equation}
$\phi_{i}$ being the orthonormalized wave functions associated
with the metal atom
, Eq. (\ref{eq:22a}) reads as follows:
\begin{equation}
   T_{ii'} =
         (V_{{\rm He}})_{ii'}
        + (E_{M}^{0}-E_{{\rm He}}^{0})
            S_{i{\rm He}} S_{{\rm He}i'} .
      \label{eq:25}
\end{equation}

Equation (\ref{eq:25}) is the fundamental equation making contact between
the repulsive potential given by Eq. (\ref{eq:11c})  and
$T_{ii'}$. Many-body contributions are partially taken into
account in Eq. (\ref{eq:25}) by means of the term
$(V_{He})_{ii'}$ which includes the bare Hartree and bare exchange
contributions,
equivalent
to $V_{{\rm electrostatic}}$  and $-J_{x}^{0} $ in Eq. (\ref{eq:14}).
The extra
term $S^{2}J_{0}$, appearing in Eq. (\ref{eq:14}) is due to the
effect of the overlap between
the $\mid {\bf k} \rangle$
and He orbitals in the total exchange interaction.

This discussion and the results of Eq. (\ref{eq:25}) suggest to
introduce
the following effective interaction between the $i$ and $i'$ orbitals
\begin{equation}
   ( T_{{\rm eff}})_{ii'} =
            S_{i{\rm He}} S_{{\rm He}i'}
         (E_{M}^{0}-E_{{\rm He}}^{0}
         -\frac{3}{8} \beta
           + \langle J^{0} \rangle ) .
          \label{eq:26a}
\end{equation}
This equation should be compared with Eq. (\ref{eq:15b})
that yields the total
repulsive potential between He and the metal atoms.

In this equation, $\langle J^{0} \rangle$ is associated with the effect of the
overlap between the He 1$s$ orbital and the atomic
wave functions of the
metal in the exchange interaction created by the He-orbital. In
Eq. (\ref{eq:15b}) ,
$J^{0}$ is the Coulomb interaction between the atomic wavefunction and
the He
1$s$ orbital; in Eq. (\ref{eq:26a}) we have introduced
$\langle J^{0} \rangle$
, the mean value of this Coulomb interaction in the crystal unit cell
(the change of $J^{0}$ along this unit cell is small, less than
10\%).

Equation (\ref{eq:26a}) is the main equation giving the effective matrix
elements
creating the excitation between the $i$ and $i'$ orbitals, or the Bloch
wave functions $\mid {\bf k} \rangle$ and  $\mid {\bf k}' \rangle$ in
the crystal, in this basis:
\begin{equation}
   ( T_{{\rm eff}})_{kk'} =
            S_{k{\rm He}} S_{{\rm He}k'}
         (E_{M}^{0}-E_{{\rm He}}^{0}
         -\frac{3}{8} \beta
           + \langle J^{0} \rangle ) .
        \label{eq:26b}
\end{equation}
\section{Dynamic interaction of Helium with a metal}
Once we have obtained the static interaction of He with the
metal, and the effective matrix elements, we will discuss
how to combine this
result with the general Eq. (\ref{eq:7}) to calculate the
stopping power for He.
First of all, let us mention that we shall use Eq.
(\ref{eq:7}) by
assuming that $\mid {\bf k} \rangle$ and $\mid {\bf k}'
\rangle$
are well described, for the He case, by
the unperturbed crystal wave functions. This is a good
approximation in our current case due to the small overlap
between the
He 1$s$ and the localized metal wave functions.

Then, the starting point is the equation
\begin{equation}
      [\hat{T}_{{\rm eff}}( {\bf R})]_{kk'} =
          V_{0} S_{k{\rm He}} S_{{\rm He}k'} ,
\end{equation}
where
\begin{equation}
          V_{0}
             =
            ( E_{M} - E_{{\rm He}}
             - \frac{3}{8} \beta
             + \langle J^{0} \rangle  ) .
\end{equation}
The overlap between the 1$s$ He state and the $\mid {\bf k} \rangle$
wave functions is written in the following way
\begin{equation}
    \langle {\bf k} \mid  \psi_{{\rm He}} \rangle =
        \int d {\bf r} \
     \psi_{k}^{*}( {\bf r}) \psi_{{\rm He}}( {\bf r})
        \simeq
     \psi_{k}^{*}( {\bf R}_{{\rm He}})
        \int d {\bf r} \
         \psi_{{\rm He}}( {\bf r}) ,
        \label{eq:28}
\end{equation}
where we replace
$
     \psi_{k}^{*}( {\bf r}) $
by
$
     \psi_{k}^{*}( {\bf R}_{{\rm He}}) $ ,
assuming the He 1$s$ level to be very localized. This allows us to
write:
\begin{eqnarray}
   \langle {\bf k}  \mid \hat{H}_{{\rm eff}}( {\bf R}) \mid
           {\bf k}' \rangle & = &
         V_{0}
           \psi_{k}^{*}( {\bf R}_{{\rm He}})
           \psi_{k'}( {\bf R}_{{\rm He}})
           [ \int d {\bf r} \
            \psi_{{\rm He}}( {\bf r}) ] ^{2}  \nonumber \\
     &  = &
        V_{0}'\psi_{k}^{*} ( {\bf R}_{{\rm He}}) \psi_{k'}
               ( {\bf R}_{{\rm He}}) .
          \label{eq:29}
\end{eqnarray}

This yields [see Eqs. (\ref{eq:7}), (\ref{eq:7a})]
\begin{eqnarray}
  H_{{\rm eff}}( {\bf q}) & = &
             V_{0}'
         \int d {\bf R}
         e^{i {\bf q} \cdot {\bf R} }
          \psi_{k}^{*} ( {\bf R}) \psi_{k'}( {\bf R}) \nonumber \\
           & = &
          V_{0}'
         I_{kk'}( {\bf q})
         \label{eq:30}
\end{eqnarray}
and
\begin{eqnarray}
  \frac{dE}{dt} & = &
        4 \pi
         \sum_{k \langle k_{F}, k' \rangle k_{F} }
        ( V_{0}' )^{2}
       \int \frac{d {\bf q} }{(2\pi)^{3}}
       \int \frac{d {\bf q}' }{(2\pi)^{3}}   \
       \frac{ ( {\bf q} \cdot {\bf v} )
              ( {\bf q'} \cdot {\bf v} )}
            { w_{kk'} }  \nonumber \\
  & & \times
       I_{kk'}( {\bf q})
        I_{k'k}( {\bf q'})
        e^{i ( {\bf q} - {\bf q'} ) \cdot {\bf R} }
        \delta ( w_{kk'} + {\bf q} \cdot {\bf v} ) .
        \label{eq:31}
\end{eqnarray}
This is the general equation giving the power loss at a given
point ${\bf R}$. Notice that
due to the crystal simmetry
\hbox {${\bf k}' = {\bf k} - {\bf q}$} and
\hbox {${\bf q}' = {\bf q} - {\bf G}$},
 ${\bf G}$ being a crystal reciprocal vector.

If we are only interested in the mean power loss and
neglect the ${\bf R}$ dependence, we should concentrate on the
${\bf q}= {\bf q'}$
contribution. Then Eq. (\ref{eq:31}) yields
\begin{eqnarray}
  \frac{dE}{dt} & = &
        4 \pi
        ( V_{0}' ) ^{2}
       \int_{-\infty}^{\infty} \frac{d {\bf k} }{(2\pi)^{3}}
       \int_{-\infty}^{\infty} \frac{d {\bf q} }{(2\pi)^{3}}   \
       ( {\bf q} \cdot {\bf v} )
       \Theta ( k_{F} - k) \Theta (k' -k_{F} ) \nonumber \\
    & & \times
        I_{kk'}( {\bf q}) \
        I_{k'k}( {\bf q}) \
        \delta ( w_{kk'} + {\bf q} \cdot {\bf v} ) ,
           \label{eq:32}
\end{eqnarray}
with
 ${\bf k'} = {\bf k} - {\bf q}$
and $\Theta$ is the step function.

Equation (\ref{eq:32}) depends on the velocity direction of the
projectile. As we shall only consider the case of He moving in alkali
metals, crystals that have a very small anisotropy, we shall calculate
the stopping power by taking an average on all the
\hbox {${\bf v}$ directions}
, which will enable us to compare our results with other works
\cite{ech:ssc37:81}. Then:
\begin{equation}
  \frac{1}{v}
  \frac{dE}{dx}  =
      \frac{
          {\normalsize \int_{-1}^{1}} d cos\theta_{v}
            dE/dt }
           {2 \ v^{2} },
        \label{eq:33}
\end{equation}
which redefines $dE/dx$.

Equation (\ref{eq:32}) is our fundamental equation for calculating the
stopping power for He, in the
low-velocity limit. This equation can be written in a local basis by
developing the ${\bf k}$ states in the atomic orbitals of the crystal.
In general, we shall assume that the metal wavefunctions are given
by an effective one-electron Hamiltonian
 $\hat{H}_{0}$, such that
\begin{equation}
  \hat{H}_{0} \mid {\bf k} \rangle = E( {\bf k}) \mid {\bf k}
       \rangle .
\end{equation}

Then, the solution of this hamiltonian yields
\begin{equation}
   \mid {\bf k} \rangle  =
   \sum_{i, \alpha}
        c_{\alpha} ( k )
        e^{i  {\bf k} \cdot {\bf R}_{i} }
        \phi_{i \alpha} ( {\bf r} - {\bf R}_{i} ) ,
          \label{eq:34}
\end{equation}
where
$\phi_{i \alpha} ( {\bf r}- {\bf R}_{i})$
are the orthonormalized wave functions  associated with the $i$
site
($
\alpha$ measures the number of orbitals per site). On the
other hand
$\phi_{i \alpha} ( {\bf r}- {\bf R}_{i})$
should be expressed as a function of the localized atomic
orbitals
$\psi_{i \alpha} ( {\bf r}- {\bf R}_{i})$
using Eq. (1). By substituting Eqs. (\ref{eq:34}) and
(\ref{eq:-1}) into Eq. (\ref{eq:32}), we find the
following result
\begin{eqnarray}
  \frac{1}{v}
  \frac{dE}{dx} &  = &
        2 \pi
       ( V_{0}' ) ^{2}
       \int_{-1}^{1} d cos\theta_{v}
       \int_{-\infty}^{\infty} \frac{d {\bf k} }{(2\pi)^{3}}
       \int_{-\infty}^{\infty} \frac{d {\bf q} }{(2\pi)^{3}}   \
       \frac{( {\bf q} \cdot {\bf v} )}{v^{2}}
       \Theta ( k_{F} - k)
       \Theta ( k' - k_{F} )  \nonumber \\
     & & \times
        \sum_{ {\bf R}_{1}, {\bf R}_{2} }
        \sum_{\alpha , \beta , \gamma , \delta }
         c_{ \alpha}^{*} ( {\bf k}) c_{ \gamma} ( {\bf k})
         c_{ \beta}^{*} ( {\bf k'}) c_{ \delta} ( {\bf k'})
                               \nonumber \\
   & & \times
        \sum_{ \alpha ' , \beta '}
        (S( {\bf k})^{-1/2})_{\alpha \alpha ' }
         I_{ \alpha ' \beta ' }^{ {\bf R}_{1} } ( {\bf q} )
        (S( {\bf k})^{-1/2})_{\beta \beta  ' }   \nonumber \\
    & & \times
        \sum_{ \gamma ' , \delta '}
        (S( {\bf k'})^{-1/2})_{\gamma \gamma ' }
         I_{ \gamma ' \delta ' }^{ {\bf R}_{2} } ( {\bf q} )
        (S( {\bf k'})^{-1/2})_{\delta \delta  ' }   \nonumber \\
    & & \times
       e^{i( {\bf k} - {\bf q}) \cdot ({\bf R}_{1}-{\bf R}_{2})}
        \delta ( w_{kk'} + {\bf q} \cdot {\bf v} ) ,
              \label{eq:35}
\end{eqnarray}
where
\begin{equation}
     I_{\beta \gamma}^{ {\bf R}_{1}} ( {\bf q}) =
         \int d {\bf r }      \
         e^{i {\bf q} \cdot {\bf r} }
          \psi_{\beta}( {\bf r})
          \psi_{\gamma}( {\bf r}- {\bf R}_{1}) ,
          \label{eq:35b}
\end{equation}
\begin{equation}
     (S( {\bf k})^{-1/2})_{\alpha \beta} =
     \sum_{ {\bf R}}
         e^{i {\bf k} \cdot {\bf R} }
       ( S^{-1/2}( {\bf R}))_{ \alpha \beta} ,
\end{equation}
\begin{equation}
     S( {\bf k})_{\alpha \beta} =
     \sum_{ {\bf R}}
         e^{i {\bf k} \cdot {\bf R} }
         \int d {\bf r}      \
          \psi_{\alpha}( {\bf r})
          \psi_{\beta}( {\bf r}- {\bf R}) .
\end{equation}

Finally, we relate
$c_{\beta}^{*} ( {\bf k}) c_{\alpha}( {\bf k})$
to the metal Green functions
$G_{\beta \alpha} ( {\bf k},w)$ by the equations
\begin{equation}
        \Theta(k_{F}-k)
         c_{\beta}( {\bf k}) c_{\alpha}^{*}( {\bf k})
         =
        \frac{1}{\pi}
          \int_{-\infty}^{E_{F}} dw \
          {\rm Im} G_{\beta \alpha} ( {\bf k},w)
\end{equation}
\begin{equation}
        \Theta(k' - k_{F})
         c_{\delta}^{*} ( {\bf k}) c_{\gamma}( {\bf k})
         =
        - \frac{1}{\pi}
          \int_{E_{F}}^{\infty} dw \
          {\rm Im} G_{\delta \gamma} ( {\bf k},w).
\end{equation}
This yields
%
\begin{eqnarray}
  \frac{1}{v}
  \frac{dE}{dx} &  = &
          2 \pi
       ( V_{0}' ) ^{2}
       \int_{-1}^{1} d cos\theta_{v}
       \int_{-\infty}^{\infty} \frac{d {\bf k} }{(2\pi)^{3}}
       \int_{-\infty}^{\infty} \frac{d {\bf q} }{(2\pi)^{3}}   \
       \frac{( {\bf q} \cdot {\bf v} )}{v^{2}}
           \nonumber \\
     & & \times
        \sum_{ {\bf R}_{1} , {\bf R}_{2} }
        \sum_{\alpha , \beta , \gamma , \delta }
         {\rm Im} [ G_{\alpha \gamma} ( {\bf k} )  ]
         {\rm Im} [ \stackrel{-}{G}_{\beta \delta} ( {\bf k} )  ]
        \nonumber \\
   & & \times
        \sum_{ \alpha ' , \beta '}
        (S( {\bf k})^{-1/2})_{\alpha \alpha ' }
         I_{ \alpha ' \beta ' }^{ {\bf R}_{1} } ( {\bf q} )
        (S( {\bf k})^{-1/2})_{\beta \beta  ' }   \nonumber \\
    & & \times
        \sum_{ \gamma ' , \delta '}
        (S( {\bf k'})^{-1/2})_{\gamma \gamma ' }
         I_{ \gamma ' \delta ' }^{ {\bf R}_{2} } ( {\bf q} )
        (S( {\bf k'})^{-1/2})_{\delta \delta  ' }   \nonumber \\
    & & \times
       e^{i( {\bf k} - {\bf q}) \cdot ( {\bf R}_{1} - {\bf R}_{2} ) }
        \delta ( w_{kk'} + {\bf q} \cdot {\bf v} ) ,
              \label{eq:38}
\end{eqnarray}
where
\begin{equation}
    G_{ \beta \alpha } ( {\bf k} ) =
                \int_{-\infty}^{E_{F}}
             \frac{dw}{\pi}
         G_{ \beta \alpha } ( {\bf k}, w)
\end{equation}
and
\begin{equation}
    \stackrel{-}{G}_{ \delta \gamma } ( {\bf k'} ) =
                \int_{\infty}^{E_{F}}
             \frac{dw'}{\pi}
         G_{ \delta \gamma } ( {\bf k'}, w') .
\end{equation}

Equation (\ref{eq:38}) allows us to calculate
the stopping power for He in
metals, as a function of the Green-function components
$G_{ \alpha \beta } ( {\bf k} )$ of the metal
(calculated in the
orthonormalized basis), using a one-electron hamiltonian
$\hat{H}_{0} ( {\bf k} )$ and the overlap matrix
$S^{-1/2} _{\alpha \beta } ( {\bf k} )$ associated with the
atomic wave functions $\psi_{\alpha}$ and $\psi_{\beta}$
. Moreover,
$ \frac{1}{v} \frac{dE}{dx} $ also depends on
$ I_{\beta \gamma} ^{ {\bf R}} ( {\bf q} ) $, the
Fourier-transform of
the overlap between the atomic orbitals
$ \psi_{\beta} ( {\bf r} ) $ and $ \psi_{\gamma} ( {\bf r} -
{\bf R} ) $ as given by Eq. (\ref{eq:35b}) .

On the other hand, in order to analyze the stopping power
as a function
of $ {\bf R} $ we take in Eq. (\ref{eq:31})
\hbox{${\bf q'} = {\bf q} - {\bf G}$}, and only the
${\bf G}$ vectors perpendicular to the ${\bf v}$ direction.
This yields for the
${\bf G}$ component of \hbox{$\frac{1}{v} \frac{dE}{dx}$},
%
\begin{eqnarray}
       S_{ {\bf G}} \equiv
       \left( \frac{1}{v} \frac{dE}{dx} \right)_{ {\bf G}}
         & = &
        4 \pi
        ( V_{0}' ) ^{2}
       \int_{-\infty}^{\infty} \frac{d {\bf k} }{(2\pi)^{3}}
       \int_{-\infty}^{\infty} \frac{d {\bf q} }{(2\pi)^{3}}   \
       \frac{( {\bf q} \cdot {\bf v} )}{v^{2}}
       \Theta ( k_{F} - k) \Theta (k' -k_{F} ) \nonumber \\
       & \  & \times
        I_{kk'}( {\bf q})
        I_{k'k}( {\bf q} - {\bf G})
        e^{ - i {\bf G} \cdot {\bf R} }
        \delta ( w_{kk'} + {\bf q} \cdot {\bf v} ) ,
           \label{eq:39}
\end{eqnarray}
and remember that \hbox{ $ {\bf k}' = {\bf k} - {\bf q} $}.

This equation  can be written in a way similar to Eq.
(\ref{eq:38}), as a function of \hbox{$S( {\bf k})$}
, \hbox{$G_{\alpha \beta} ( {\bf k} )$} and
, \hbox{$ I_{\alpha \beta} ^{{\bf R}}$}
. For the sake of brevity, we
shall only mention here that  in general the stopping power
\hbox{$S = \frac{1}{v} \frac{dE}{dx}$}
can be written as follows:
%
\begin{equation}
   S_{ {\bf R} } = S_{0} +
       \sum_{ {\bf G}}
          S_{ {\bf G}} e^{i {\bf G} \cdot {\bf R} } ,
      \label{eq:40:II}
\end{equation}
as a function of ${\bf R}$,
where $S_{0}$ is the mean stopping power given by Eq.
(\ref{eq:38}), and $S_{ {\bf G}}$ the ${\bf G}$ component
of Eq.
(\ref{eq:39}).
Once we have chosen the ${\bf G}$ vectors perpendicular to
${\bf v}$, we have calculated $S_{ {\bf G}}$ by taking
an average on the angle between ${\bf v}$ and ${\bf q}$
as in Eq. (\ref{eq:33}).
\section{Results and discussions}
We have applied the previous formalism to the calculation of the
stopping
power for He in alkali metals. For simplicity, the band is assumed
to be
well described by means of a single $s$ orbital. Then, Eq.
(\ref{eq:35}) can be further
simplified into the following equation:
\begin{eqnarray}
  \frac{1}{v}
  \frac{dE}{dx} &  = &
         2 \pi
        ( V_{0}')^{2}
       \int_{-1}^{1} d cos\theta_{v}
       \int_{-\infty}^{\infty} \frac{d {\bf k} }{(2\pi)^{3}}
       \int_{-\infty}^{\infty} \frac{d {\bf q} }{(2\pi)^{3}}   \
       \frac{( {\bf q} \cdot {\bf v} )}{v^{2}}
       \Theta ( k_{F} - k)
       \Theta ( k' - k_{F} ) \nonumber \\
     & & \times
        \sum_{ {\bf R}_{1} , {\bf R}_{2} }
        (S( {\bf k})^{-1/2})
                       I^{ {\bf R}_{1}}( {\bf q})
             (S( {\bf k})^{-1/2})
        (S( {\bf k'})^{-1/2})
                       I^{ {\bf R}_{2}}( {\bf q})
             (S( {\bf k'})^{-1/2}) \nonumber \\
     & & \times
       e^{i( {\bf k} - {\bf q}) ( {\bf R}_{1} - {\bf R}_{2} ) }
        \delta ( w_{kk'} + {\bf q} \cdot {\bf v} ) ,
        \label{eq:40:IV}
\end{eqnarray}
or
\begin{eqnarray}
  \frac{1}{v}
  \frac{dE}{dx} &  = &
         2 \pi
        ( V_{0}' )^{2}
       \int_{-1}^{1} d cos\theta_{v}
       \int_{-\infty}^{\infty} \frac{d {\bf k} }{(2\pi)^{3}}
       \int_{-\infty}^{\infty} \frac{d {\bf q} }{(2\pi)^{3}}   \
       \frac{({\bf q} \cdot {\bf v} )}{v^{2}}
       \Theta ( k_{F} - k)
       \Theta ( k' - k_{F} ) \nonumber \\
     & & \times
        \sum_{ {\bf R}_{1} , {\bf R}_{2} }
        S( {\bf k})^{-1}
        S( {\bf k'})^{-1}
      \left(    \sum_{ {\bf R}_{1}}
              e^{i ({\bf k}-{\bf q}) \cdot {\bf R}_{1}}
              I^{{\bf R}_{1}}( {\bf q})
          \right)
      \left(    \sum_{ {\bf R}_{2}}
              e^{-i ( {\bf k}-{\bf q}) \cdot {\bf R}_{2}}
              I^{{\bf R}_{2}}({\bf q})
          \right)       \nonumber       \\
      & & \times
        \delta ( w_{kk'} + {\bf q} \cdot {\bf v} ) ,
         \label{eq:40}
\end{eqnarray}
where
\begin{equation}
     I^{ {\bf R}_{1}} ({\bf q}) =
         \int d {\bf r}      \
         e^{i {\bf q} \cdot {\bf r} }
          \psi( {\bf r})
          \psi( {\bf r}-{\bf R}_{1})
         \label{eq:41a}
\end{equation}
and
\begin{equation}
     S( {\bf k}) =
     \sum_{ {\bf R}}
         e^{i {\bf k} \cdot {\bf R} }
          \int d {\bf r} \
          \psi( {\bf r})
          \psi( {\bf r}- {\bf R}) .
\end{equation}
Equation (\ref{eq:41a}) can be written in a more
symmetric way as follows. Take
\begin{eqnarray}
         {I}^{ {\bf R}_{1}} ({\bf q}) & = &
         e^{i {\bf q} \cdot {\bf R}_{1}/2 }
         \int d {\bf r}      \
         e^{i {\bf q} \cdot ( {\bf r}-{\bf R}_{1}/2 ) }
          \psi({\bf r})
          \psi({\bf r}-{\bf R}_{1})    \nonumber \\
         & =  &
       e^{i {\bf q} \cdot {\bf R}_{1} /2 }
          \stackrel{-}{I} ^{{\bf R}_{1}} ({\bf q}),
\end{eqnarray}
then
\begin{equation}
      \sum_{{\bf R}_{1}}
          e^{i ({\bf k}-{\bf q}) \cdot {\bf R}_{1}}
           I^{{\bf R}_{1}}({\bf q})   =
      \sum_{{\bf R}_{1}}
          e^{i ({\bf k}-{\bf q}/2) \cdot {\bf R}_{1}}
          \stackrel{-}{I} \ ^{{\bf R}_{1}}({\bf q})
         \label{eq:43a}
\end{equation}
and
\begin{equation}
      \sum_{{\bf R}_{2}}
          e^{-i ({\bf k}-{\bf q}) \cdot {\bf R}_{2}}
           I^{{\bf R}_{2}}({\bf q})   =
      \sum_{{\bf R}_{2}}
          e^{-i ({\bf k}-{\bf q}/2) \cdot {\bf R}_{2}}
          \stackrel{-}{I} \ ^{{\bf R}_{2}}({\bf q}) .
        \label{eq:43b}
\end{equation}

Equations (\ref{eq:40:IV}) and
(\ref{eq:40})
 yield the stopping power for  He as a function
of $S({\bf k})$ and
      $\stackrel{-}{I} \ ^{{\bf R}}({\bf q})$.
$S({\bf k})$  has been calculated using the atomic
wave functions given in Ref. \cite{cle:andt14:74}
. The calculation of
      $    \stackrel{-}{I} \ ^{{\bf R}}({\bf q})    $
is more complicated since the Fouriertransform of the
atomic
wavefuncions centered on different sites are needed.
This is the well-known problem of
multicenter integrals. Several solutions have been tried in
the literature
 \cite{wor:ss258:91} such as expanding the
Slater-type basis in a
Gaussian one  \cite{mor:nim:92,boy:prsa200:50} .
We have used, however,
an adaptative algorithm
of integration by Monte Carlo techniques \cite{pet:jcp27:78} to
perform \hbox{$\stackrel{-}{I} \ ^{{\bf R}}$}.

An approximate solution, which yields good results for $S_{0}$,
is
obtained by replacing
\begin{equation}
          \stackrel{-}{I} \ ^{ {\bf R} }( {\bf q})
                    \simeq
          S( {\bf R} ) \ *  \  I( {\bf q} ) ,
        \label{eq:44}
\end{equation}
where
\begin{equation}
          S( {\bf R} ) =
          \int d {\bf r} \
          \psi( {\bf r})
          \psi( {\bf r}- {\bf R})
\end{equation}
and
\begin{equation}
          I( {\bf q} ) =
          \int d {\bf r} \
         e^{i {\bf q} \cdot  {\bf r} }
          \psi({\bf r})
          \psi({\bf r}) .
\end{equation}

Equation (\ref{eq:44}) is exact in the limit
${\bf R}_{i} = 0$ or ${\bf q} \rightarrow 0$.
In general, we expect Eq. (\ref{eq:44}) to give
a good approximation
to \hbox{$I^{{\bf R}_{i} } ({\bf q})$}
if  $ {\bf q} \cdot {\bf R}_{i} /2$ is small.

Introducing Eq. (\ref{eq:44}) into Eq.
(\ref{eq:40}) yields:
\begin{eqnarray}
  \frac{1}{v}
  \frac{dE}{dx} &  = &
          2 \pi
        ( V_{0}'  )^{2}
       \int_{-1}^{1} d cos\theta_{v}
       \int_{-\infty}^{\infty} \frac{d {\bf k} }{(2\pi)^{3}}
       \int_{-\infty}^{\infty} \frac{d {\bf q} }{(2\pi)^{3}}   \
       \frac{({\bf q} \cdot {\bf v} )}{v^{2}}
       \Theta ( k_{F} - k)
       \Theta ( k' - k_{F} ) \nonumber \\
     & \times &
        \frac{I(q)}{S( {\bf k})}
        \frac{I(q)}{S({\bf k'})}
        \mid
           \sum_{{\bf R}}
                   S({\bf R})
              e^{i ({\bf k}-{\bf q}/2) \cdot {\bf R} }
        \mid ^{2}  \
        \delta ( w_{kk'} + {\bf q} \cdot {\bf v} ) .
         \label{eq:45a}
\end{eqnarray}

Equation (\ref{eq:45a}) is the basis of our approximation
to Eqs.
(\ref{eq:40:IV}) and
(\ref{eq:40}). We
should also mention that $ E_{k} $  (the electron energy band of
the
alkali metal) has been assumed to follow a free electron
dispersion law.

Before discussing the numerical results given by
Eq. (\ref{eq:45a}), it is
worth considering the results obtained by neglecting all
the overlaps
between the alkali atom wavefunctions. Then we write
\begin{eqnarray}
     S( {\bf R}) =
         \left \{
           \begin{array}{cc}
                1,  &  \ \ {\bf R} = 0 \\
                0,  &  \ \ {\bf R} \neq 0 \\
           \end{array}
         \right.
\end{eqnarray}
and
\begin{equation}
     S({\bf k}) =  1
\end{equation}
and replace Eq. (\ref{eq:45a}) by the following equation:
\begin{eqnarray}
  \frac{1}{v}
  \frac{dE}{dx} &  = &
         2 \pi
       ( V_{o}' )^{2}
       \int_{-1}^{1} d cos\theta_{v}
       \int_{-\infty}^{\infty} \frac{d {\bf k} }{(2\pi)^{3}}
       \int_{-\infty}^{\infty} \frac{d {\bf q} }{(2\pi)^{3}}   \
       \frac{({\bf q} \cdot {\bf v} )}{v^{2}}
       \Theta ( k_{F} - k)
       \Theta ( k' - k_{F} ) \nonumber \\
     & \times &
        \left|
         \int d {\bf r} \
         \psi^{2}({\bf r})
         e^{i {\bf q} \cdot {\bf r}  }
        \right| ^{2}
        \delta ( w_{kk'} + {\bf q} \cdot {\bf v} ) .
         \label{eq:46}
\end{eqnarray}

It is also convenient to discuss at this point the stopping power
given
by the following simple model: a uniform electron gas interacting
with a slowly  moving He atom by means of the following contact
potential
\begin{equation}
    \hat{H} _{{\rm pert}} =
          V_{0}'
          \delta ( {\bf r} - {\bf v} t )
        \label{eq:47} .
\end{equation}
Here $V_{0}'$ is assumed to be the same local potential
introduced in
Eq. (\ref{eq:29}). It
is an easy task to develop this model following the same steps
as discussed above for the LCAO approach and find the following
expression
for the stopping power
\begin{equation}
  \frac{1}{v}
  \frac{dE}{dx}  =
        4 \pi
       ( V_{0}')^{2}
       \int_{-\infty}^{\infty} \frac{d {\bf k} }{(2\pi)^{3}}
       \int_{-\infty}^{\infty} \frac{d {\bf q} }{(2\pi)^{3}}   \
       \frac{({\bf q} \cdot {\bf v} )}{v^{2}}
       \Theta ( k_{F} - k)
       \Theta ( k' - k_{F} )
        \delta ( w_{kk'} + {\bf q} \cdot {\bf v} ).
         \label{eq:48}
\end{equation}
The integral in ${\rm cos}\theta_{v}$ equals 2 because in the latter
expression
$  \frac{1}{v}
  \frac{dE}{dx} $
depends only on
$ \mid {\bf v} \mid $ .

Comparing Eqs. (\ref{eq:46}) and (\ref{eq:48}),
we see that their only difference
is associated  with the term
  \hbox {$
          \mid I(q) \mid ^{2} =
           \mid
           \int d {\bf r} \
           \psi ^{2} ( {\bf r})
           e^{i {\bf q} \cdot {\bf r}  }
          \mid ^{2} $
          }, which
gives the form factor of the metal orbital.
We should also comment, regarding Eq. (\ref{eq:45a})
that in the alkali metals
 \hbox {$
           S({\bf k}) \sim S(k_{F}) $
         }
since, in the low velocity limit we are considering $ {\bf k} $
and
${\bf k'}$ are located near the Fermi sphere,
\hbox{$ S({\bf k}) $}
being almost constant on this surface that presents a very small
anisotropy. Then,
Eq. (\ref{eq:45a}) can be obtained from Eq.
(\ref{eq:46}) by
replacing the form factor \hbox{ $ I ({\bf q}) $} by
\begin{equation}
  D( {\bf q}, {\bf k} ) =
            \frac{I(q) S( {\bf k}- {\bf q} /2 ) }
                { S({\bf k}) } ,
        \label{eq:49}
\end{equation}
where
\begin{equation}
      S( {\bf k}- {\bf q}/2) =
         \sum_{ {\bf R}}
         S( {\bf R})
         e^{i ( {\bf k} -{\bf q} / 2) \cdot {\bf R} } .
\end{equation}

Thus the three different cases we are considering yield the same
equation for the stopping power, but for a specific factor taking
the
values 1, $I(q)$ and $ D({\bf q}, {\bf k} )$, for the
free-electron
gas (FEG), the
LCAO model with $ S ({\bf R}) = 0 $ for $ {\bf R} \neq 0 $
(LCAO-I), and the
LCAO model with $ S ({\bf R}) \neq 0 $  (LCAO-II),
respectively. What is of interest to realize
about this discussion is that the free-electron-gas model
overestimates
the stopping power, while the simplest LCAO model underestimates
it. In
Table \ref{tab1}, we give the three values of the mean stopping power
\hbox{$S_{0}$}, for He
in Na  as calculated from these equations.
As shown in this Table \ref{tab1}
the free electron gas model yields a
stopping
power three times too large, while in the LCAO-I model $
\frac{1}{v} \frac{dE}{dx}$ is about  eight times too
small.

One word of caution must be put here. The FEG model discussed
here can
not be compared directly with the
 LDA used to
calculate the stopping power of He in metals. The point to notice
is that in the model defined by Eq. (\ref{eq:47}),
$V_{0}'$
is the contact potential for the interaction of He with the
$s$ orbitals
of the alkali-metal atoms. The model of Eq. (\ref{eq:47}) is only
introduced here
in order to explain how the form factor of Eqs. (\ref{eq:49})
or (\ref{eq:46}) is the
main term controlling the He stopping power.

As regards the factor
$D({\bf q}, {\bf k} )$
used to calculate
$\frac{1}{v}
  \frac{dE}{dx}$
in the LCAO-II approximation, notice the strong
dependence that
$D({\bf q}, {\bf k} )$
has on the number of neighbors used to calculate
\hbox{
   $   S({\bf k} ) =
      \sum_{{\bf R}}
       S({\bf R})
       e^{i {\bf k} \cdot {\bf R} }   $
   }
and
   $  S( {\bf k}-{\bf q}/2 )      $
in Eq. (\ref{eq:49}). We have found that in order to get a
reasonable accuracy (around \hbox{5\%}) it is necessary to
add up to the fifth or sixth
neighbor, depending on the alkali metal.

As mentioned above, Eq. (\ref{eq:40}) has been accurately
calculated for Na using Monte Carlo
techniques. We have found that this Monte Carlo calculation
yields \begin{equation}
  \frac{1}{v}
  \frac{dE}{dx}  =     0.085 \ a.u.  \ ({\rm Na}),
\end{equation}
a value a little larger than the one found using our LCAO-II
approximation. By assuming the same correction factor for all the
alkali
metals, we find the results given in Table \ref{tab2}, column
(a). This Table
also
shows the theoretical figures obtained by Echenique, Nieminen,
and Ritchie
\cite{ech:ssc37:81}.

 We see from Table \ref{tab2}, column (a),
 that the results for K and Rb are in
excellent agreement with Ref. \cite{ech:ssc37:81},
although the stopping powers we find for Li and Na are a little
larger.
This difference can be partially attributed to the simple model we are
using, since
a single $s$ orbital per alkali-metal atom has been assumed to form
the metal conduction band. This approximation can be expected to be
a reasonable one for very electropositive atoms like K and Rb
, but not so appropriate at least for Li.
 Thus, in the calculations of Papaconstantopoulos
\cite{pap:hbses:86} for Li, only
52\% of the occupied density of states has a $s$ like character.
If we
introduce in the results of Table II, a factor
\begin{equation}
       n_{s}^{2} / n_{s}^{2} \ ({\rm Rb} )
\end{equation}
which normalizes the stopping power of each alkali metal to the
total
number of $s$ electrons with respect to Rb, we find the
results of Table
\ref{tab2}, column (b), in much better
agreement with the LDA calculations.

The conclusion we can draw from these results is that the method
developed in this paper is quite appropriate to calculate the
stopping
power for He moving slowly in alkali metals. We can also expect
that the
method will be useful to calculate stopping powers for atoms in
transition metals.

In a further step we have calculated, using
Monte Carlo techniques,
the stopping power dependence on the ion position
(for He moving in a
channeled direction). We have considered that
He moves in a Na crystal
along the [100] direction. We have calculated the different
${\bf G}$ reciprocal vectors  contributing to
the stopping power [Eq. (\ref{eq:39})]; this implies taking
the $
{\bf G}$ vectors perpendicular to the [100] direction. In a bcc
lattice, the first reciprocal vectors to be considered are the
followings: \hbox{${\bf G} \equiv \frac{2 \pi}{a} (0,1,1) $,}
      \hbox{$ \frac{2 \pi}{a} (0,0,2) $} etc.  Using Eqs.
(\ref{eq:39}) and (\ref{eq:40:II}) we have obtained the
stopping power Fourier components \hbox{$ S_{G} $}
shown in Table \ref{tab4}.

Figure 3 shows \hbox{$ S_{{\bf R}} $}, with ${\bf R}$
changing in a
[100] plane. The main conclusion we can draw from these
calculations is
the strong dependence that the stopping power shows as a
function of the
impact parameter: the stopping power can vary as much as
\hbox{100\%} for
different impact distances. We should comment that these
changes are not
associated with the electronic metal charge
\cite{gra:pla:92}; this charge,
as obtained in our LCAO approach
with an $s$ level per atom, appears to be almost constant in the
crystal lattice except very close to the atomic sites.
For a He-atom channeled along the Na [100] direction,
one expects some kind of oscillatory motion of the atom,
with the impact parameter changing along
the He trajectory. Then, the mean-stopping power for the
channeled case
would appear as an average of the different values shown
in Fig. 3 around
the minimum value of the stopping power. Each case should
be analyzed
specifically, but assuming the incoming atom to explore only
half of the
total available space, one would get around 50-60\% of $S_{0}$,
namely 0.04 a.u., 80\% of the value calculated in LDA.
\section{Conclusions}
The aim of this work has been to develop a first-principles,
free-parameter, approach based on a LCAO method to calculate the
stopping
power for atoms
 moving in condensed matter. In the past few years the interest
in,
generally speaking, tight-binding methods \cite{har:esps:80}
for analyzing the
electronic properties of solids has increased a lot. This
emphasis is
partially due to the interest in using a local point of view,
closely
related to the chemistry of the local environment. The work presented
in this
paper follows this general
trend and tries to apply the ideas recently developed in Refs.
\cite{gol:prb39:89,gar:prb44:91}
for analyzing the electronic properties of solids following a
LCAO method,
to the stopping power area. In the long term, this approach can
be expected to be also useful for analyzing other dynamical
processes like
the charge transfer between moving ions and the solid, sticking
mechanisms, etc.

In Sec. II, we have presented our general approach and have
related
the stopping power for atoms, in the low-velocity limit, to the
electronic properties of the crystal as described using a LCAO
method.
All the parameters appearing in Eq. (\ref{eq:7}), the general
equation giving
the stopping power, can be obtained from the local wave funcions
of the
atoms forming the crystal. Equation (\ref{eq:7}) has been applied
to the case of
He moving in alkali metals. He is a simple atom, but the alkali
metals
present a strong test to our method as their atomic wave funcions
interact
strongly with each other up to large separations.
In Sec. IV, we have presented
our results and have found that the stopping power for  He is
very well described with our local LCAO approach, if the
interaction between different alkali metal atomic orbitals
is included, at least, up to fifth neighbor.

We conclude that the LCAO method discussed in this paper
offers the
possibility of calculating accurately the stopping power for ions
moving
in solids. This could be a convenient framework for analyzing
solids having
localized $d$ bands and for discussing specific geometries like the
case of
atoms moving near surfaces, or the channeled case discussed in
Sec. V.

\acknowledgments
This work has been partially funded by the Spanish CICYT under
contract
no. PB89-165. One of the authors (J.J.D.) thanks Ministerio de
Educaci\'on y Ciencia and  Universidad Aut\'onoma de Madrid,
Spain,
for their financial support. F.F. acknowledges support by
Iberdrola
S.A. We thank R. Ritchie, N. Lorente, P.M. Echenique, and M.
Jakas  for helpful discussion.

\begin{figure}
\caption{He moving inside a metal.}
     \label{fig1}
\end{figure}

\begin{figure}
\caption{Schematic representation of a He atom interacting with a metal
        band simulated by a metal level $E_{M}$.}
      \label{fig2}
\end{figure}

\begin{figure}
\caption{ Stopping power for He moving inside a Na crystal along
the [100]-direction. The stopping power is normalized with respect to
its mean value. The coordinates correspond to a (100) plane
perpendicular to the projectile trajectory. The atomic rows
along the [100] direction are
projected onto the points having the coordinates: (0,0), (0,1), (1,1),
(1,0) and, (0.5,0.5).}
 \label{fig3}
\end{figure}

\begin{table}
  \begin{tabular}{llll}
       \/       & FEG    &   LCAO-I    & LCAO-II  \\ \hline
  $  \frac{1}{v}
     \frac{dE}{dx} $ (a.u.)  &     0.140 & 0.007 & 0.056 \\
  \end{tabular}
\caption{
Stopping power for He in Na as calculated for the free electron
model (FEG), the LCAO model neglecting the metal wave functions overlaps
(LCAO-I), and the
LCAO model taking into account these overlaps (LCAO-II).
        }
    \label{tab1}
\end{table}

\begin{table}
\begin{tabular}{llll}
  \/ & (a)  & (b)    & (c) \\  \hline
Li & 0.260  & 0.110  & 0.100 \\
Na & 0.085  & 0.068  & 0.053 \\
K & 0.026   & 0.023  & 0.023 \\
Rb & 0.015  & 0.015  & 0.016 \\
\end{tabular}
\caption{
        Stopping power in a.u. for He in different alkali metals.
        (a) Our results, as calculated using Monte Carlo
techniques.
        (b) Our results with the $s$ ocupancy correction.
        (c) ENR, from Ref. \protect\cite{ech:ssc37:81}. }
    \label{tab2}
\end{table}


\begin{table}
\begin{tabular}{lr}
  \/      & atomic units      \\  \hline
S(0,0,0)  & \hbox{ $ 0.85 \times  10^{-1} $ } \\
S(0,1,1)  & \hbox{ $ 0.18 \times 10^{-1} $ }  \\
S(0,0,2)  & \hbox{ $ -0.04 \times 10^{-1} $ } \\
\end{tabular}
\caption{ Stopping Power for He in Na for the
          channeled direction [100]
        \label{tab4}
        }
\end{table}

\begin{references}
\bibitem{rut:pm21:11}
    E. Rutheford, Philos Mag. {\bf 21}, 669 (1911).

    N. Bohr, {\em ibid.} {\bf 25},16 (1913).
\bibitem{bet:ap5:30}
    H.A. Bethe, Ann. Phys. (Leipzig) {\bf 5}, 325 (1930).
\bibitem{fer:zf29:27}
    E. Fermi, Z. Phys. {\bf 29}, 315 (1927).
\bibitem{wil:rmp17:45}
    E.J. Williams, Rev. Mod. Phys. {\bf 17}, 217 (1945).
\bibitem{lin:mfm28:54}
    J. Lindhard, K. Dans, Mat. Fys. Met. {\bf 28}, 8 (1954).
\bibitem{bra:acs:75}
    W. Brant, in {\em Atomic Collisions in Solids } , edited by
   S. Datz, B.R. Appelton,
and C.D. Moak (Plenum, New York, 1975).
\bibitem{rit:jpcs26:65}
    R.H. Ritchie and J.C. Ashley, J. Phys. Chem. Solids {\bf 26}
, 1689 (1965).
\bibitem{arn:prl65:90}
    A. Arnau, M. Pe\~{n}alba, P.M. Echenique, F. Flores, and
R.H. Ritchie, Phy. Rev. Lett. {\bf 65}, 1024 (1990).
\bibitem{flo:icpss:91}
   F. Flores, in {\em Interaction of Charged Particles with Solids and
Surfaces}, edited by  A. Gras-Marti {\em et al. } (Plenum, New York, 1991).
\bibitem{ech:ssps43:90}
   P.M. Echenique, F. Flores, and R.H. Ritchie, Solid State Physics
Series {\bf 43}, 229 (1990).
\bibitem{ech:ssc37:81}
   P.M. Echenique, R.M. Nieminen, and R.H. Ritchie, Solid State Commun.
{\bf 37}, 779 (1981).
\bibitem{bau:prl69:92}
   P. Bauer, F. Kastner, A. Arnau, A. Salin, P.D. Fainstein,
V.H. Ponce, and P.M. Echenique, Phys. Rev. Lett. {\bf 69}, 1137
(1992).
\bibitem{gra:pla:92}
       P.L. Grande and G. Schwietz, Phys. Lett. A {\bf 163}, 439 (1992).
\bibitem{gol:prb39:89}
   E.C. Goldberg, A. Martin-Rodero, R. Monreal, and F. Flores, Phys Rev.
B {\bf 39},5684 (1989).
\bibitem{gar:prb44:91}
   F.J. Garc\'{\i}a Vidal, A. Martin-Rodero, F. Flores, J. Ortega, and
R. Perez
, Phys. Rev. B {\bf 44},
11412
(1991).
\bibitem{sab:pra42:90}
   J.R. Sabin, J. Oddershede, and G.H.F. Diercksen, Phys. Rev. A
{\bf 42}, 1302 (1990).

   D.E. Meltzer, J.R. Sabin, and S.B. Trickey, {\em ibidi.}
{\bf 41}, 220 (1990).
\bibitem{Low}
   P.O. L\"{o}wdin, J. Chem. Phys. {\bf 18}, 365 (1950).
\bibitem{sol:td:85}
   F. Sols, Ph.D. Thesis, Universidad Aut\'onoma de Madrid (1985).

   F. Sols and F. Flores, Solid State Commun. {\bf 42}, 687 (1982).
\bibitem{cle:andt14:74}
   E. Clementi and C. Roeti, {\em Atomic Data and Nuclear Tables}
{\bf 14}, 177 (1974).
\bibitem{wor:ss258:91}
     H. Wormeester, H.J. Borg, and A. Van Silfhout, Surf. Sci.
{\bf 258}, 197 (1991).
\bibitem{mor:nim:92}
  E.H. Morknersen, J. Oddershede and J.R. Sabin, Nucl. Instrum.
Methods Phys. Rev. (to be published).
\bibitem{boy:prsa200:50}
        S.F. Boys, Proc. R. Soc. London Ser A {\bf 200}, 542 (1950).
\bibitem{pet:jcp27:78}
   G. Peter Lepage, J. of Comput. Phys. {\bf 27},
192 (1978).
\bibitem{pap:hbses:86}
   D.A. Papaconstantopoulos, in {\em Handbook of Band Structure
of the Elemental Solids} (Plenum, New York, 1986).
\bibitem{har:esps:80}
   W. A. Harrison, in {\em Electronic Structure and Properties
of Solids }
(Freeman, San Francisco, 1980).
   A.D. Sutton, M.W. Finnis, D.G. Pettiford, and Y. Ohta, J. Phys. C
{\bf 21}, 35 (1988);
  J.A. Majewshi, and P. Vogl, Phys. Rev. Lett. {\bf 57}, 1366 (1986).
\end{references}
\end{document}